# Sensors Lifetime Enhancement Techniques in Wireless Sensor Networks - A Survey

Dr. Sami Halawani, Abdul Waheed Khan

**Abstract -** Wireless Sensor Networks are basically used for gathering information needed by smart environments but they are particularly useful in unattended situations where terrain, climate and other environmental constraints may hinder in the deployment of wired/conventional networks. Unlike traditional networks, these sensor networks do not have a continuous power supply at their disposal. Rather the individual sensors are battery operated and the lifetime of the individual sensors and thus the overall network depend heavily on duty cycle of these sensors. Analysis on WSNs shows that communication module is the main part which consumes most of the sensor energy and that is why energy conservation is the major optimization goal. Since routing protocols and MAC protocols directly access the communication module therefore the design of protocols in these two domains should take into account the energy conservation goal. In this paper, we discuss different state-of-the-art protocols both in MAC and routing domains that have been proposed for WSNs to achieve the overall goal of prolonging the network lifetime. The routing protocols in WSNs are generally categorized into three groups – data centric, hierarchical and location-based but we focus on only the first two categories because location-based routing protocols generally require a prior knowledge about sensors location which most of the times is not available due to random deployment of the sensors. We then discuss how schedule-based and contention-based MAC protocols can contribute to achieve optimal utilization of the limited energy resource by avoiding or reducing the chances of collisions and thus the need for retransmission.

- - - - - - - - - - - - - - - - - - - ◆ - - - - - - - - - - - - - - - - - - - -

## 1 INTRODUCTION

W<small>IRELESS</small> Sensor Network (WSN) is a network of large numbers – up to thousands – of tiny spatially distributed radio-equipped sensors. Each node in a sensor network is composed of a radio-transducer, a small microcontroller and a long lasting battery for energy source. These sensor networks are used for gathering information needed by smart environments and are particularly useful in unattended situations where terrain, climate and other environmental constraints may hinder in the deployment of wired/conventional networks. An individual node failure is not an issue because of the large scale deployment of these nodes and normally the target area is monitored by several nodes. Primarily these sensors are used for data acquisition and are required to disseminate the acquired parameters to special nodes called sinks or base-stations over the wireless link as shown in figure 1. The base-station or sink collects data from all the nodes, and then analyzes this data to draw conclusions about the on-going activity in the area of interest [1]. Sinks or base- stations being powerful data processors can act as gateways to other existing communications infrastructure or to the Internet where a user can have access to the reported data.

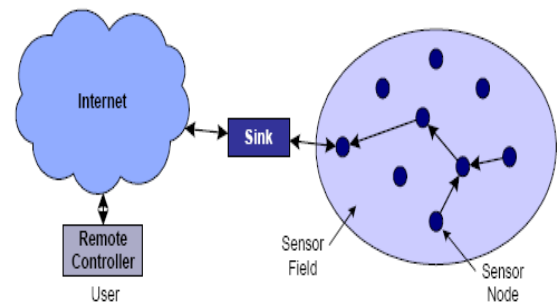

Fig. 1. Sensor Network Architecture

In this paper, we first describe what is main optimization goal of WSNs? We then describe in detail what efforts have been made with regards to achieve energy conservation in WSNs in section 3, where we essentially discuss routing and media access protocols for WSNs. Finally, section 4 concludes our discussion and points out some future innovations in this field.

## 2 PROBLEM STATEMENT

Analysis on WSNs shows that communication module is the main part which consumes most of the sensor energy and that is why energy conservation is the major optimization goal. Keeping in mind the energy constraints combined with a typical deployment of large number of sensor nodes pose many challenges to the design and management of WSNs and necessitate energy-awareness at all layers of the networking protocol stack [2]. In addition to routing protocols, the Medium-Access-Control (MAC) protocols directly control the communication module so protocols in these two domains need to be reshaped in such a way to optimally utilize the energy resource.

## 3 METHODOLOGY

As communication module is the major consumer of the sensor energy resource and the routing protocols and MAC protocols directly control the communication module so the protocols in these two domains need to be refined according to the energy constraints of WSNs. So we will focus on different protocols working in these two domains to efficiently utilize the energy resource and thus achieve the main design goal i.e., prolong the lifetime of the sensor networks.

### 3.1 Energy Conservation using Routing Protocols

Routing in wireless sensor networks is very challenging due to the inherent characteristics that distinguish these networks from other wireless networks [2] for several reasons: Firstly, the global addressing scheme does not fit well in WSNs due to relatively large number of sensors and accordingly high cost of ID maintenance. Secondly, unlike the traditional IP networks, most of the times the sensors are deployed in ad hoc manner where the sensors act as self-organized entities and work cooperatively with each other without being controlled by some central devices. Finally, a target area in WSNs is usually monitored by several sensors and thus there is a high probability that the data that they transmit has some redundancy. Thus to improve energy and bandwidth utilization, this redundancy needs to be tackled by some intelligent routing protocols.

Many new algorithms have been proposed for the routing problem in WSNs which take into consideration the inherent features of WSNs along with the application and architecture requirements. Node lifetime shows a strong dependence on the battery lifetime [3]. In a multihop WSN since there is no dedicated access point so each node plays a dual role as data sender and data router. The malfunctioning of some sensor nodes due to power failure can cause both frequent topological updates and rerouting of lost packets and that would shorten the lifetime of the network thereby eating up the energy. Hence, creation of routing tables is both insignificant and impractical. To achieve efficient utilization of the energy resource many routing techniques have been proposed in the literature which employ concepts like data aggregation and in-network processing, clustering, different node role assignment, and data-centric methods and the list goes on. Almost all of the routing protocols for WSNs can be classified as data-centric, hierarchical or location-based. Before discussing these different categories of protocols, first let us take into account the system architecture design issues for sensor networks and their implications on data routing because these routing protocols have taken into consideration the inherent features of WSNs along with the application and architecture requirements [2].

### 3.1.1 Architectural Design Issues in WSNs

The primary design goal of WSNs is to acquire the monitored data from target area and deliver it to base-station for its evaluation, while trying to prolong the lifetime of the network. The design of routing protocols in WSNs is influenced by many challenging factors and these factors must be considered to achieve efficient communication in WSNs. In the following section, we briefly discuss some of these design issues that affect routing process in WSNs.

### a. Data Aggregation

To ensure full coverage of the target area, multiple sensors are deployed to monitor the area so that in case of individual node failures, the area is still in coverage of other sensors. However, this also results in significantly redundant data generated by these sensors which need to be tackled by some mechanism to reduce the number of transmissions and thus energy consumption there by employing some aggregate function on this acquired data. Data aggregation is the combination of data from different sources by using functions such as suppression (eliminating duplicates), min, max and average [4]. Since computation consumes less energy than communication so great energy savings can be obtained through data aggregation. This technique has been used to achieve energy efficiency and traffic optimization in a number of routing protocols [3][5]. In some network architectures, all aggregation functions are assigned to more powerful and specialized nodes [6]. Signal processing methods can also be used for data aggregation. In that case, it is referred to as data fusion where a node is capable of producing a more accurate output signal by using some techniques such as beamforming to combine the incoming signals and reducing the noise in these signals [7].

### b. Node Deployment

Nodes deployment can be deterministic or randomized also known as self-organized and nodes deployment is application specific. In deterministic deployment, sensors are manually placed and data is routed through pre-determined paths. Deterministic deployment is less power hungry because routes are already known and incase an event occurs, it does not require to determine routes thereby saving energy. In random node deployment, the sensor nodes are scattered randomly creating an infrastructure in an ad hoc manner [7] and if the nodes distribution is not uniform, optimal clustering is employed to allow connectivity and achieve efficient network operation. However, inter-nodes distance should be kept shorter usually a few meters keeping in mind the energy and bandwidth limitation because larger distances consume more energy. Therefore in WSNs, most of the times multi-hop architecture is employed.

### c. Energy Model

To transmit a signal over a distance r, the required radiation energy is proportional to $r^m$ where m is 2 in the free space and ranges up to 4 in environments with multiple-path interferences or local noise.

Therefore during the creation of a wireless network infrastructure, the process of setting up the routes is greatly influenced by energy considerations. Since even in ideal environments, the transmission power of a wireless radio is proportional to distance squared, hence in most of environments multihop routing is employed because that shortens the inter-sensors distance and thus consumes less energy than direct communication. However, multi-hop routing introduces significant overhead for topology management and medium access control. Direct routing would perform well enough if all the nodes are very close to the sink [7].

### d. Data Delivery Model

Data sensing and delivery in WSNs is application dependent and the data delivery model can be continuous (time-driven), event-driven, query-driven and hybrid [8]. In the continuous delivery model, each sensor sends data periodically. The time-driven delivery model is suitable for applications that require periodic data monitoring. The nodes will periodically switch on their sensors and transmitters, sense the environment and transmit the acquired data at constant periodic time intervals and then again go to sleep mode. In event-driven and query-driven models, sensor nodes react immediately to sudden and drastic changes in the value of a sensed attribute due to the occurrence of a certain event or a query is generated by the sink. As such, these are well suited for time critical applications. A combination of these different models is also possible. Therefore, the choice of routing protocol is highly influenced by the specific application environment and thus the data delivery model with regard to energy consumption.

### e. Node Capabilities

Sensor nodes can be homogeneous or heterogeneous. In homogeneous environment, all nodes have equal capacity in terms of computation, communication and power and all of them have the same role i.e., sensing the data, performing some lightweight computation and then transmitting it over the wireless channel to other nodes or sink. They can also work as router and reroute the received data to others. However, depending on the application, a sensor node can have dedicated role or functionality such as relaying, sensing and aggregation since incorporating the three functionalities at the same time in a single node might quickly consume the energy of that node. Some of the hierarchical protocols proposed in the literature designate a cluster-head different from the normal sensors [2]. Most researchers propose that a cluster-head should be more powerful than the sensor nodes in terms of energy, bandwidth and memory and the burden of transmission to the sink and aggregation tasks should be handled by the cluster-head. However some pick the cluster-heads from the deployed sensors because the heterogeneous set of sensors raises multiple technical issues related to data routing [9].

In the following section, we discuss various classical mechanisms and state-of-the-art routing protocols specifically designed for WSNs. Since location-based routing protocols generally require a prior knowledge about sensors location which most of the times is not available due to random deployment of the sensors. So we will limit our discussion to protocols working in data-centric and hierarchical families of protocols along with their potentials of how much energy conservation they can achieve. We will also discuss which protocols are suited best for which applications and scenarios.

### 3.1.2 Data-Centric Routing Protocols

In most of the applications of sensor networks, the sensors are densely deployed to cope with frequent failures of individual sensors and to ensure uninterrupted coverage of the target area. In such kind of scenarios a multihop routing mechanism is employed to efficiently utilize the energy resource and sensor nodes are typically of homogeneous nature thereby playing the same role of sensing and routing the data in a collaborative manner. Since these nodes are usually densely deployed so assigning a global identifier to each node is not a feasible solution and the base-station has no direct means to select a specific set of sensor nodes to be queried. Therefore, data is usually transmitted from every sensor node within the deployment region with significant redundancy [2]. However, this is very inefficient in terms of energy consumption because the same duplicate information is routed by multiple sensors and therefore the routing protocols should be able to select a set of sensor nodes which can perform data aggregation during relaying of data. This consideration has led to data centric routing where the base-station sends queries to certain regions and waits for the response from the sensors located in the selected regions [10]. Since data is being requested through queries, attribute-based naming is necessary to specify the properties of data. SPIN [11] and Directed Diffusion [3] are among the first proposed data-centric routing protocols which consider data negotiation between nodes in order to eliminate redundant data and save energy.

In this section, we describe some classical mechanisms, SPIN and Directed Diffusion routing protocols, followed by some new protocols that employ the concepts of SPIN and Directed Diffusion.

### a. Flooding and Gossiping

Flooding and gossiping [12] are the two classical mechanisms that have been borrowed from traditional communication network to relay data in sensor networks without the need for any routing algorithms and topology maintenance. In flooding, each sensor node receives a data packet and broadcasts it to all of its neighbors where the neighbors further broadcast it to their neighbors and this process continues until the packet arrives at the destination or the maximum number of hops for the packet is reached. Flooding is a very easy mechanism but it has several drawbacks as shown in figures 2 and 3.

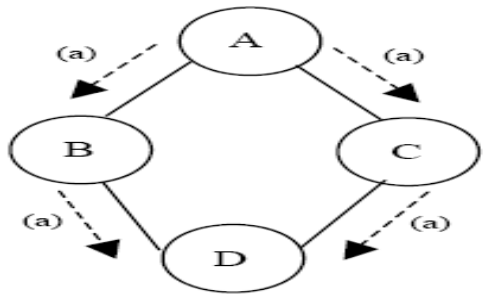

Fig. 2. The implosion problem. Node A starts by flooding its data to all of its neighbors. D gets two same copies of data eventually, which is not necessary.

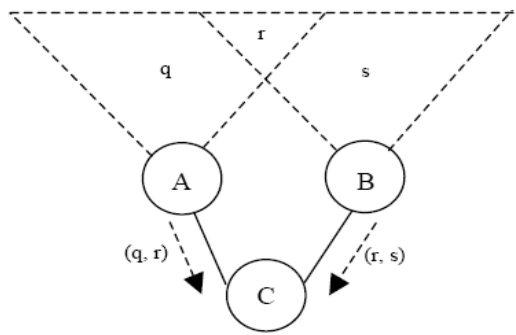

Fig. 3. The overlap problem. Two sensors A & B cover an overlapping geographic region and C gets same copy of data from these sensors.

'Implosion' and 'Overlap' are the two main drawbacks of flooding which waste the energy resource. 'Implosion' is caused by duplicate messages sent to the same node and 'Overlap' is caused when two nodes sensing the same region will send similar packets to the same neighbor.

Gossiping is an enhanced version of flooding which avoids the problem of implosion by just selecting a random node to send the packet to rather than broadcasting the packet blindly which picks another random neighbor to forward the packet to and the process goes on. However, this cause delays in propagation of data through the nodes.

### b. Sensor Protocol for Information via Negotiation (SPIN)

SPIN is a family of adaptive protocols proposed by Heinzelman et al, in [11] that disseminate all the information at each node to every node in the network assuming that all nodes in the network are potential base-stations. SPIN uses data negotiation and resource-adaptive algorithms. Nodes running SPIN assign a high-level name to completely describe their collected data (called meta-data) and perform meta-data negotiations before any data is transmitted. Using this meta-data negotiation mechanism SPIN solves the classic problems of flooding such as redundant information passing and overlapping of sensing areas and saves considerable amount of energy resource. SPIN does not enforce any specific meta-data format and the format is assumed application specific. In addition, SPIN has access to the current energy level of the node and adapts the protocol it is running based on how much energy is remaining. To exchange data between nodes SPIN defines just three messages ADV, REQ and DATA. ADV message is used to allow a sensor to advertise a particular meta-data, REQ message is used to request the specific or advertised data and DATA message represents the actual data that is sent upon request. The protocol starts when a SPIN node obtains new data that it is willing to share. It does so by broadcasting an ADV message containing meta-data. If a neighbor is interested in the data, it sends a REQ message for the DATA and the DATA is sent to this neighbor node. The neighbor sensor node then repeats this process with its neighbors. As a result, the entire sensor area will receive a copy of the data. Figure 4 summarizes the steps of the SPIN protocol.

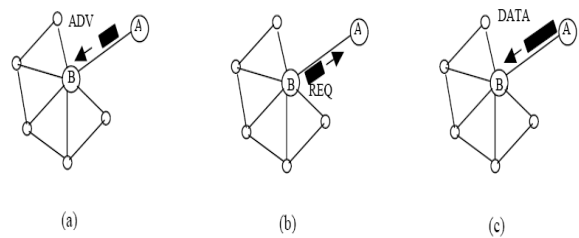

Fig. 4. SPIN Protocol. (a) Node A starts by advertising its data to node B. (b) Node B responds by sending a request to node A. (c) Node A then transmits actual data to node B.

The SPIN family of protocols includes many protocols and the two main protocols are called SPIN-1 and SPIN-2, which incorporate negotiation before transmitting data in order to ensure that only useful information is transferred. The SPIN-1 is a 3-stage process as discussed above. SPIN-2 is an extension to SPIN-1 and it incorporates an additional threshold-based resource awareness mechanism. In case when energy is abundant in sensors, SPIN-2 works the same way as SPIN-1 but when the energy starts falling below a certain threshold, it reduces participation in the protocol meaning it participates only when it believes that it can complete all the other stages of the protocol.

The main advantage of SPIN is that topological changes are localized since each node needs to know only its single-hop neighbors. SPIN greatly overcomes the shortcomings of flooding and using the meta-data negotiation mechanism almost halves the redundant data. However, SPINs data advertisement mechanism cannot guarantee the delivery of data. For instance, if the source and destination nodes are far away from each other nodes that are interested in the data are far away from the source node and the nodes between source and destination are not interested in that data, such data will not be delivered to the destination at all. That is why, SPIN is not a good choice for applications such as intrusion detection, which require reliable delivery of data packets over regular intervals.

**c. Directed Diffusion**

Directed Diffusion is one of the first proposed query-driven data-centric routing protocols for WSNs proposed by Intanagonwiwat et al, in [3]. This protocol uses simple attribute-based naming as the fundamental building block. Both requests for information (called interests) and the notifications of observed events are described through sets of attribute–value pairs. The main idea of the Directed Diffusion paradigm is to combine the data coming from different sources enroute (in-network aggregation) by eliminating redundancy, minimizing the number of transmissions and thus saving network energy and prolonging its lifetime.

The steps of the basic Directed Diffusion are as follows and the whole operation can be viewed as in figure 5:

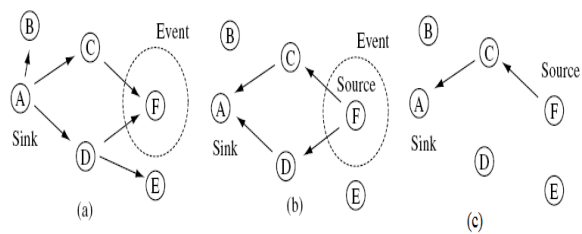

Fig. 5. Directed Diffusion Routing

(i) A sink initiates an interest for a specific type of information by flooding it throughout the network (the overhead of this can be reduced if necessary by using geographic scoping or some other optimization). The interest may be periodically repeated if robustness is called for.

(ii) Every node in the network caches the interest while it is valid, and creates a local gradient entry towards the neighboring node(s) from which it heard the interest. The sink's ID/network address is not available and hence not recorded, however the local neighbors are assumed to be uniquely identifiable through some link-layer address. The gradient also specifies a value (which could be an event rate, for instance).

(iii) A node which obtains sensor data that matches the interest begins sending its data to all neighbors it has gradients toward. If the gradient values stand for event rates then the rate to each neighbor must satisfy the gradients on the respective link. All received data are cached in intermediate nodes to prevent routing loops.

(iv) Once the sink starts receiving response data to its interest from multiple neighbors, it begins reinforcing one particular neighbor (or k neighbors, in case multi-path routing is desired), requesting it to increase the gradient value (event rate). These reinforcements are propagated hop by hop back to the source. The determination of which neighbor to reinforce can take into account other considerations such as delay, link quality, etc. Nodes continue to send data along the outgoing gradients, depending on their values.

(v) (Optional) Negative reinforcements are used for adaptability. If a reinforced link is no longer useful/efficient, then negative reinforcements are sent to reduce the gradient (rate) on that link. The negative reinforcements could be implemented by timing out existing gradients, or by re-sending interests with a lower gradient value.

Essentially what Directed Diffusion does is (a) the sink lets all nodes in the network know what the sink is looking for, (b) those with corresponding data respond by sending their information through multiple paths, and (c) these are pruned via reinforcement so that an efficient routing path is obtained.

There are two major differences between Directed Diffusion and SPIN. In Directed Diffusion, the sink queries the sensor nodes if a specific data is available by flooding some tasks. In SPIN, sensors advertise the availability of data allowing interested nodes to query that data. Second, all communication in Directed Diffusion is neighbor-to-neighbor with each node having the capability of performing data aggregation and caching in addition to sensing. Caching is a big advantage in terms of energy efficiency and delay. Unlike SPIN, there is no need to maintain global network topology in Directed Diffusion. However, due to query-driven nature of Directed Diffusion, it cannot be applied to all sensor network applications that require continuous data delivery to the base-station e.g., environmental monitoring. This is because the query-driven on-demand data model may not help in this regard. Also matching data to queries might require some extra overhead at the sensor nodes.

### d. Rumor Routing

Rumor routing proposed by Braginsky et al, in [13] is a variation of Directed Diffusion and is applicable for applications where geographic routing is not feasible. Generally Directed Diffusion uses flooding to inject the query to the entire network incase if there is no geographic criterion to diffuse tasks. However sometimes only a little amount of data is requested from the nodes and in such cases the use of flooding is unnecessary. Rumor routing employs an alternative approach according to which instead of flooding the query into the network, flood the events if the number of events is small and the number of queries is large. The key idea is to route the queries to only those nodes who have observed a particular event rather than flooding the entire network to inquire about an event. For this purpose, the rumor routing algorithm employs long-lived packets, called agents. When a node detects an event, it adds that event to its local table, called events table, and generates an agent. Agents travel through the network in order to propagate information about local events to distant nodes. Whenever a node generates a query for an event, the nodes that know the route, may respond to the query by inspecting its event table. Therefore there is no need to flood the whole network and thus reducing the number of transmissions.

Simulation results have shown that rumor routing achieves significant energy saving compared to event-flooding [2]. Rumor routing is useful only when the number of events is small. For a large number of events, the cost of maintaining agents and event-tables in each node becomes infeasible if there is not enough interest in those events from the base-station.

### e. Energy Aware Routing (EAR) Protocol

EAR proposed by Shah et al. in [14] is another data-centric routing protocol which is somehow similar to Directed Diffusion. It differs from Directed Diffusion in the sense that it maintains a set of paths instead of maintaining or enforcing one optimal path at higher rates. The approach argues that using the minimum energy path all the time will deplete the energy of nodes on that path. Instead, one of the multiple paths is used with a certain probability so that the whole network lifetime increases. Thus by efficient load-balancing mechanism, the energy of any single path will not deplete quickly. This approach prolongs the network lifetime as energy is dissipated more equally among all nodes. These paths are chosen by means of a probability function, which depends on the energy consumption of each path. EAR is comprised of three phases:

- **Setup Phase**
  Localized flooding is performed to find the routes and create the routing tables. Also in this process, the total energy cost is calculated in each node. Paths with very high cost are discarded. The node selection is done according to closeness to the destination. The node assigns a probability to each of its neighbors in forwarding table which is inversely proportional to the cost.
- **Data Communication Phase**
  Using the probability distributions, each node forwards the packet by randomly picking a node from its forwarding table.
- **Route Maintenance Phase**
  Localized flooding is performed infrequently to keep all the paths alive.

EAR and Directed Diffusion mechanisms differ in only one way i.e., in Directed Diffusion data is sent through multiple paths and then one of them is being reinforced to send at higher rates. While EAR selects a single path randomly from the multiple alternatives in order to save energy. Using this random path selection mechanism, EAR achieves an overall improvement 21.5% in energy saving compared to Directed Diffusion [2]. On the negative side, does not provide any mechanism to recover from a node or path failure as opposed to Directed Diffusion.

### 3.1.3 Hierarchical Routing Protocols

A flat network also known as single-tier (sensors to sink) network may potentially degrade the network performance in case of large scale deployment of sensors. To allow WSNs to deal with a large scale deployment of sensors and thus cover a large area of interest, hierarchical routing is employed that divides the entire network into multiple clusters. The objective of hierarchical routing is to manage the energy consumption of WSNs by setting up multihop communication within a particular cluster and by performing data aggregation and data fusion at the cluster-heads to decrease the number of transmission to the sink. The cluster-heads could be chosen from ordinary sensors. However, in a typical hierarchical architecture, heterogeneous sensors are used where higher energy nodes are selected as cluster-heads and are assigned the tasks of obtaining data from other sensors, performing aggregation and then relaying towards the sink while low energy nodes are assigned the sensing tasks in the proximity of the target. This tasks distribution mechanism greatly contributes to overall system scalability and energy efficiency.

In this category of routing protocols, we discuss LEACH, PEGASIS, TEEN and APTEEN.

### a. Low-Energy Adaptive Clustering Hierarchy (LEACH) Protocol

LEACH proposed by Heinzelman et al, in [7] is one of the early and most popular hierarchical routing protocols for sensor networks. It forms clusters of the sensor nodes based on the received signal strength and then randomly selects some sensors as the cluster-heads. These local cluster-heads are used as routers to the sink. LEACH rotates the role of cluster-head to balance energy dissipation of sensors. The decision of which node will become the cluster-head is made by the node choosing a random number, r, between 0 and 1. The node becomes a cluster-head for the current round if the number is less than the following threshold value T(n):

$$T(n) = \begin{cases} \dfrac{p}{1 - p * (r \bmod \frac{1}{p})} & \text{if } n \in G \\ 0 & \text{otherwise} \end{cases}$$

where p is the desired percentage of cluster heads, r is the current round, and G is the set of nodes that have not been cluster heads in the last 1/p rounds.

The cluster-head extracts sensed data from local sensors and after performing aggregation sends the aggregated data to the sink. In this way, LEACH reduces both the amount of information and the number of transmissions that must be forwarded to the sink.

Using the clusters and data aggregation mechanism, LEACH achieves over a factor of 7 reduction in energy dissipation compared to direct communication [2]. Another advantage of LEACH is that it is completely distributed and requires no global knowledge of network. However, it is not applicable to networks deployed in large regions because it uses single-hop routing in a cluster where each node can transmit directly to the cluster-head and the sink. Similarly, the formation of dynamic clustering pose extra overhead, e.g. head changes, advertisements etc., which may consume extra energy.

### b. Power Efficient in Sensor Information System (PEGASIS) & Hierarchical-PEGASIS

PEGASIS proposed by Lindsey et al, in [5] is a variation of LEACH protocol. Unlike LEACH, PEGASIS avoids multiple cluster formation and forms chains of sensor nodes and only one node is selected from that chain as a chain-leader to transmit to the base-station. Sensed data moves from node to node, gets aggregated by each node and is finally sent to the base-station. As shown in Figure 6 node n0 passes its data to node n1. Node n1 aggregates node n0's data with its own and then transmits to the chain-leader i.e., n2. After node n2 passes the control to node n4, node n4 transmits its data to node n3. Node n3 aggregates node n4's data with its own and then transmits to the chain-leader n2. Node n2 waits to receive data from both neighbors and then aggregates its data with its neighbors' data. Eventually, node n2 transmits this message to the base-station.

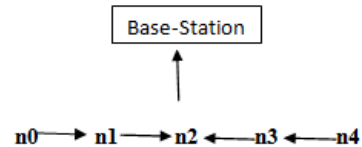

Fig. 6. Chaining Operation in PEGASIS

In PEGASIS each node uses the signal strength to locate the closest neighbor node by measuring the distance to all neighboring nodes and then adjusting the signal strength so that only one node can be heard. This chain will consist of those nodes that are closest to each other and form a path to the base-station.

PEGASIS has been shown to increase the lifetime of the network twice as much the lifetime of the network under the LEACH protocol for different network sizes and topologies [2]. This mainly stems from the elimination of the overhead caused by dynamic cluster formation in LEACH and through decreasing the number of transmissions and reception by using data aggregation at multiple levels. However, the data collection mechanism by chain-leader from its neighbors incurs excessive delay for distant node on the chain. PEGASIS although avoids the clustering overhead of LEACH yet they still require dynamic topology adjustment because sensor energy is not tracked. For instance, every sensor must be aware of its neighbor status so that it knows where to route its data. This topology adjustment again incurs overhead especially for highly utilized networks.

Hierarchical-PEGASIS proposed by Lindsey et al, in [15] is a modified version of PEGASIS, which tries to reduce the delay that occurs during transmission of packets to the base-station and proposes a solution to the data gathering problem by considering energy×delay metric. To reduce the delay, Hierarchical-PEGASIS pursues simultaneous transmissions of data messages. However, simultaneous transmissions may result in collisions due to shared medium. To avoid such collisions and possible signal interference among the sensors, two approaches are adopted. The first approach incorporates signal coding, e.g. CDMA. In the second approach simultaneous transmissions occur only when the sensors are spatially far apart from each other. The chain-based protocol with CDMA capable nodes, constructs a chain of nodes, that forms a tree like hierarchy, and each selected node in a particular level transmits data to the node in the upper level of the hierarchy. This method reduces the delay considerably by ensuring data transmission in parallel. Using the second approach a three-level hierarchy of the nodes is formed and the interference effects during simultaneous transmission are reduced by carefully scheduling the transmissions. It has been shown that this chain-based protocol performs better than the regular PEGASIS scheme by a factor of about 60 [10].

### c. Threshold-sensitive Energy Efficient sensor Network (TEEN) & Adaptive Periodic TEEN (APTEEN) Protocols

TEEN is a hierarchical protocol proposed by Manjeshwar et al, in [16]. This protocol was proposed for time-critical applications in order to be responsive to sudden changes in the sensed attributes. For time critical applications such as fire detection in forests, responsiveness is very important and the network should operate in a reactive mode. TEEN uses a hierarchical approach and it also employs data-centric mechanism. In TEEN, the sensor network architecture is based on a hierarchical grouping however unlike LEACH, it forms multi-level clusters where closer nodes to the target form clusters and this process goes on to the second level until the base-station is reached. This model is shown in figure 7.

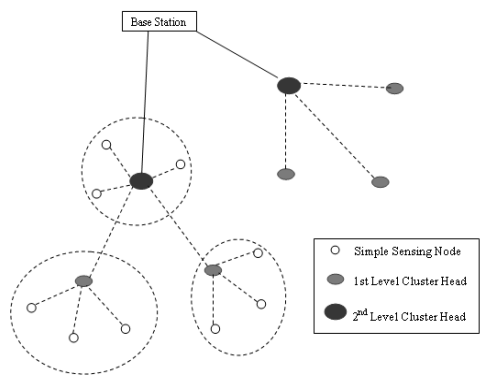

Fig. 7. Hierarchical Clustering in TEEN

After forming the clusters, each cluster-head broadcasts two thresholds to the nodes. One is hard threshold and the other is soft threshold. The hard threshold that is broadcasted by cluster-head to the cluster nodes is minimum possible value of an attribute to trigger a sensor node to switch on its transmitter and transmit to the cluster-head. So the hard threshold mechanism allows the nodes to transmit only when the sensed attribute is in the range of interest and in this way it reduces the number of transmissions significantly.

Upon sensing a value at or beyond the hard threshold, the node transmits the sensed data only if the value of that attribute changes by an amount equal to or greater than the soft threshold. As a result, soft threshold further reduces the number of transmissions if there is no significant change in the value of sensed attribute. Further the values of both hard and soft thresholds are adjustable so one can easily control the number of packet transmissions according to the application requirements. For example, a smaller value of the soft threshold gives a more accurate picture of the network, at the expense of increased energy consumption. So there is a trade-off between energy efficiency and data accuracy and this needs to be taken into account.

Since TEEN is a kind of reactive routing protocols so it is suitable only for time-critical applications. For applications where periodic reports are needed, TEEN may not be a good choice because the user may not get any data at all if the thresholds are not reached.

APTEEN proposed by Manjeshwar et al, in [17] is an enhanced version of TEEN which aims to make TEEN versatile thereby capturing periodic data collections as well as reacting to time-critical events. APTEEN follows the same clustering philosophy of TEEN, however in APTEEN the cluster-head also performs data aggregation for the sake of saving energy. Furthermore, APTEEN supports three different query types: historical – to analyze the past data values; one-time – to take a snapshot view of the network; and persistent – to monitor an event for a period of time.

Experiment studies show that APTEEN's performance lies between LEACH and TEEN in terms of energy consumption [2]. Both TEEN & APTEEN perform very well with regards to energy consumption because they decrease the number of transmissions. However the overhead and complexity associated with formation of multi-level clusters hinders in its wide scale adaptability. Also the threshold mechanism and dealing with attribute-based naming of queries limit their use to some specific applications.

## 3.2 Energy Conservation using MAC Protocols

In wireless networks, communication occurs in the form of electromagnetic signal transmission in free air. This differs from wired networks which employ guided medium for communication. This common transmission medium must therefore be shared by all sensor network nodes in a fair manner. To achieve this goal, a medium access control protocol is employed. The choice of the medium access control protocol is the major determining factor in WSN performance. To cope with the energy limitation of wireless sensors, the MAC protocols for wireless sensor networks are refined to accommodate an additional goal of managing radio activity to optimally utilize the energy resource [18]. Energy efficiency is the key requirement in the design of MAC protocol for wireless sensor nodes. There are several factors that lead to energy inefficiency in WSNs [19]. The most profound source of energy waste is 'collision' which occurs when two or more nodes simultaneously attempt to transmit data. Consequently, to retransmit the data corrupted by a collision increases energy consumption. Another artifact in traditional MAC protocols that leads to energy waste is 'idle listening'. A node is said to be in 'idle listening' mode when it is listening for a traffic that is not sent. The third source of energy waste is 'overhearing' which occurs when a sensor node receives packets that are destined to other nodes. The fourth major source of energy waste is caused by control packet overhead. Control packets are required to regulate access to the transmission channel. A high number of control packets transmitted, relative to the number of data packets delivered indicates low energy efficiency. Finally, frequent switching between different operation modes may result in significant energy consumption. Limiting the number of transitions between sleep and active modes, for example, leads to considerable energy saving [19].

Different MAC protocols for WSNS have been proposed in literature with the main objective to reduce energy waste caused by collisions, idle listening, overhearing and excessive overhead. Broadly, they can be divided into two main groups [19] – Schedule-based and Contention-based MAC Protocols. In the following section, we discuss each of these categories of protocols and the mechanisms that they employ to avoid or reduce collisions and thus the need for retransmissions.

### 3.2.1 Schedule-Based MAC Protocols

In WSNs to allow sensors to gain access to the shared wireless medium in a cooperative manner, schedule-based MAC protocols have been proposed that regulate access to resources according to a schedule to avoid contention among nodes. Depending upon the medium access technique, the resources could be time slot, frequency band or a CDMA code. The main aim of schedule-base MAC protocols is to achieve a high degree of energy consumption to prolong the network lifetime. Most of the schedule-based MAC protocols for WSNs use a variant of a Time-Division-Multiple-Access (TDMA) scheme whereby the channel is divided into time slots [19] as shown in figure 8. Using this scheme, a logical frame of N contiguous slots is formed and this logical frame repeats itself in cycle over time. Each sensor node is assigned a set of specific time slots per frame and this set constitutes the schedule according to which the sensor node gains access to the medium and have the right to transmit or receive. This schedule can be either fixed or constructed on demand on a per-frame basis by the base-station to reflect the current requirements of sensor nodes and traffic pattern, or it could be hybrid. The nodes must also satisfy the interference constraint which says that no nodes within two hops of each other may use the same slot. This two-hop constraint is needed to avoid hidden-node problem that is discussed later on. Energy conservation is achieved by using an on/off mechanism of the sensor's radio transceiver. According to the schedule of each sensor node, a sensor alternates between two modes of operation i.e., active mode and sleep mode. A sensor is said in active mode when it is its turn to use the assigned time-slots within the logical frame to transmit and receive data frames. Outside a sensor's assigned time-slots, it moves into sleep mode by switching off its radio transceiver.

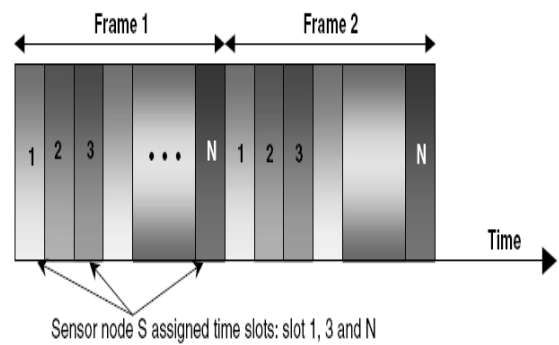

Fig. 8. TDMA based MAC Protocols for WSN

Several versions of the basic TDMA protocol have been proposed for media access control in WSNs. In the following section, we briefly discuss these protocols with regards to achieve energy conservation in WSNs.

### a. Self Organizing Medium Access Control for Sensor Networks (SMACS)

In TDMA-based MAC protocols, the time slots assignments to nodes can be performed either in a centralized manner or a decentralized manner. The centralized manner requires that a base-station gathers the full network topology, perform the slot assignments offline and then distribute it back to the network. However, such solutions are not effective if network size grows and particularly with dynamic environments [18]. Distributed approaches are therefore employed for this purpose. The SMACS protocol proposed by Sohrabi et al, in [20] is based on distributed approach which enables nodes to discover their neighbors then establish communication schedules without the assistance of any local or global master node. In essence, during the startup phase, each node decides on a common communication slot with its neighbor through handshaking on a common control channel. Each communication link consists of a pair of time slots operating at a randomly chosen, but fixed frequency (or CDMA code). In WSNs since available bandwidth can be expected to be much higher than the maximum data rate for sensor nodes, so it is assumed that there are sufficiently many frequencies/codes to ensure that there are no common frequency/time assignments within interference range, and hence there is no contention. Energy conservation is achieved by using a random wake-up schedule during the connection phase and by turning the radio off during idle time slots.

### b. Energy Efficient MAC Protocol for Sensor Networks (EMACS)

EMACS is also a TDMA-based MAC protocol proposed by Dulman et al, in [21]. As described in [21], EMACS divides the time into so called frames where each frame is divided into time slots and each slot further contains three sections: Communication Request (CR), Traffic Control (TC), and data section. There is no sharing of a time slot and one time slot belongs to just a single node. The owner of the time slot is the decision maker who tells its neighbors to whom it wants to communicate during its time slots. Other nodes may demand some data or inform about the data that they possess to the owner of the time slot in the CR section. The owner transmits its schedule for its data section by broadcasting a table in the TC section and after the TC section, the actual transmission of data packets follows. As soon as a node finishes its transmission or receiving operation, it should turn off its radio transceiver (go to sleep mode) to conserve energy. The TC section can also be used to make wakeup calls for sleeping nodes. EMACS introduces two sleep modes for nodes:

- **Standby mode**
  A node goes into standby mode when no transmissions are expected at a certain time. Consequently, the node releases its time slots and to keep in touch with the network it then periodically listens to a TC section of a frame. In event-driven and query-driven applications, if the node has to react and transmit some data, it simply fills up the CR section of another node and agrees on the data transmission, completes the transmission and then go back to sleep mode.
- **Dormant mode**
  In this mode, the sensors adapt a low power mode for an agreed amount of time and this time is set by coordination of higher layers. The sensor then wakes up, performs synchronizations to rediscover the network and then starts communication based on its assigned time slots. This sleep mode is particularly useful to tackle redundancy in the network.

EMACS reduces the utilization of transceivers of the nodes to save energy as much as possible. It also reduces latency in the network by releasing the owned time slots in one of the sleep modes. However a drawback of not owning a time slot is that the nodes will not be able to receive unicast messages that are directly addressed to it and it will only receive multicast or broadcast messages.

### c. Distributed Energy Aware MAC (DE-MAC) Protocol

DE-MAC protocol proposed by Kalidindi et al, in [22] is a robust MAC protocol for WSNs in the sense that sensor nodes not only access the medium according to a schedule but they also take into account the relative energy levels of neighbor nodes. Like other schedule-based MAC protocols, DE-MAC also employs the periodic listen and sleep mechanism to avoid idle-listening and overhearing. However, unlike other schedule-based MAC protocols, it treats critical nodes (nodes left with short energy) differently in a distributed fashion. The design of DE-MAC is inspired by the fact that over a period of time, some nodes may deplete there energy quickly than others and may die earlier. Consequently, the average useful life of the sensor network is reduced. To prolong the lifetime of these critical sensors and thus the overall sensor network, DE-MAC proposes that these critical nodes must be treated differently with respect to energy consumption. DE-MAC initially assigns the same number of transmission slots to each node in a TDMA frame. To determine the critical sensors, a local election is conducted that is based on energy of neighboring nodes. The local election process does not result in any throughput loss because this process is fully integrated with regular TDMA communication schedule. The election process is initiated by any node whose current energy level falls below a threshold value of the previous winner's then-energy level. As soon as an election is initiated, each node transmits special energy-level messages that are appended to its regularly scheduled transmission packet during its assigned time slot. Another distinction of DE-MAC from other TDMA protocols is that all nodes listen to all transmitted packets and there are no sleeping nodes when other nodes are transmitting. Finally, at the end of the election process, the node with least energy among the participants is elected as leader. As a result, all losers reduce their number of slots by a constant factor (typically 2) and the winners get time slots twice that of losers. The aim behind this election process and reallocation of slots is to reduce the idle listening time of those critical nodes. Nodes also turn off their radio and go to sleep node when they have nothing to transmit during their assigned slots thus completely eliminating idle listening.

There are two main advantages of DE-MAC [22]: First since DE-MAC is a schedule-based MAC protocol, therefore it is collision free and thus conserves energy by avoiding collisions and thus retransmission of packets. Second since slots are pre-allocated to each node so it does not require any contention mechanism. However the key feature of DE-MAC is the consideration of relative energy levels of sensors and the fair election process for the leader to make efficient utilization of energy resource of critical sensors.

### 2.2.2 Contention-Based MAC Protocols

Contention-based MAC protocols are also known as Random-Access MAC protocols. These protocols do not pre-allocate resources to individual sensors. Instead, they employ an on-demand channel access mechanism and in this way share a single radio channel among the contending nodes. Simultaneous attempts to access the communications medium, however, results in collision. Effectively, these protocols try to minimize rather than completely avoid the occurrence of collisions.

Traditional networks use CSMA (Carrier Sense Multiple Access) as a medium access mechanism. However, CSMA mechanism gives poor performance in WSNs due to two unique problems: the hidden node problem and the exposed node problem. The hidden node problem is shown in figure 9 (a) where node A is transmitting to node B and node C which is out of coverage of A will sense the channel as idle and start packet transmission to node B as well. Consequently, the two packets will collide at node B. In this case CSMA fails to foresee this collision. The exposed node problem is shown in figure 9 (b), where while node B is transmitting to node A, node C has a packet intended for node D. Since node C is in range of B, it will sense the channel as busy and will postpone its transmission. However, in theory, D is outside the range of B, and A is outside the range of C and therefore these two transmissions would not collide with each other.

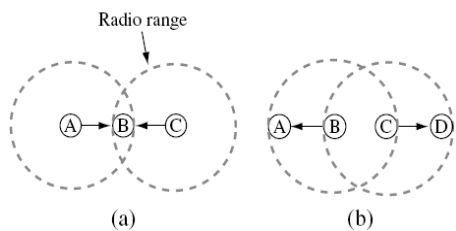

Fig. 9. Problems with basic CSMA in wireless environment (a) hidden node (b) exposed node

Several protocols have been proposed in the literature that aim at minimizing the chances of collisions and follow a mechanism called Carrier-Sense-Multiple-Access with Collision Avoidance (CSMA/CA).

### a. Medium Access with Collision Avoidance (MACA)

The MACA protocol proposed by Kam in [23], makes use of two control messages – request to send (RTS) and clear to send (CTS), to avoid collisions. According to MACA, when a node wishes to send a message, it first sends an RTS message to its intended recipient and if the recipient is willing to receive, it issues a CTS message to the sender. Hence, when the sender receives the CTS, it begins to transmit the actual data packet. Similarly, if a nearby node listens an RTS meant for another node, it restrains its own transmission for a while and waits for a CTS response. If a CTS message is received, this node restrains its own transmission for a sufficient time to allow the other parties to complete their communication and in this way avoids collision.

Under ideal conditions that no collisions of RTS and CTS messages will occur, MACA performs very well with regards to avoid collisions and thus reduces the number of retransmissions. Using the RTS and CTS control messaging mechanism, MACA effectively solves both the hidden node problem and the exposed node problem. However the wait & watch mechanism incurs some delay.

### b. Power Aware Medium Access with Signaling (PAMAS)

PAMAS technique introduced by Singh et al, in [24] is an improved version of MACA technique which employs two separate radio channels – one for RTS/CTS signaling and another for data exchange. PAMAS in one of the power aware MAC protocols proposed for multihop wireless networks. In PAMAS, sensor nodes switch off their radio and go to sleep mode whenever they can neither transmit nor receive successfully. Node goes to sleep mode whenever it detects a neighbor transmitting to another node or when it determines through the control channel RTS/CTS signaling that one of its neighbors is receiving. The duration of the sleep mode is set to the length of the ongoing transmissions which is indicated by the control signals received on the secondary channel. If a node is in sleep mode and in the mean-time another transmission starts, then upon wakeup the node sends probe signals to determine the duration of the ongoing transmission and again goes to sleep mode for the remaining time.

PAMAS saves considerable amount of energy by switching off its radio when it detects that other parties are transmitting. However, when the node is in idle reception mode i.e., when it has no packets to send and there is no activity on the channel, a considerable amount of energy goes into waste.

### c. Sensor MAC (S-MAC)

S-MAC proposed by Yee et al, in [25] is a robust MAC protocol for WSNs that achieves significant reduction in energy consumption. S-MAC specifically aims at reducing energy consumption in listening to idle channel by incorporating a periodic sleep mechanism. S-MAC combines the features of both contention-based and schedule-based MAC protocols in the sense that the channel access mechanism is contention-based but the sleep-wakeup intervals are schedule-based. The nodes conserve energy by going into sleep mode (switching off its radio) and then waking up periodically to listen and check if any of its neighbors are interested in communication with it. The 'duty cycle' of a node is the ratio of the time it is awake (i.e. not in the sleep state) to the total time [26] and the lower the duty cycle, the lower is the energy consumption of a sensor node. In sensor networks, all nodes may not have the same duty cycle and some nodes may deplete their energy faster than others. This reduces the average life of the WSN nodes, and hence reduces the average useful life of the sensor network itself. So a schedule-based mechanism may further prolong lifetime of the sensor network.

The listening and sleeping intervals are of fixed length and are set according to application requirements. Each frame is comprised of a listen interval followed by a sleep interval. If a node wants to send data to another node, it must initiate the communication process during the listening interval of its intended receiver and other nodes at this stage go back to their sleep mode and thus avoid overhearing and idle listening. However since nodes periodically sleep so an interested sending node must know the listening and sleeping schedule of a neighbor with which it wants to communicate. In order to have this knowledge, nodes exchange their schedules by periodic broadcasts of a special control frame called 'sync frame' and there is subinterval reserved for this control frame in the listen interval in each frame. The sync control frames are broadcasted through CSMA/CA protocol. The nodes need to broadcast the sync frame at least once in a predetermined synchronization period. Accordingly each node builds a table of the schedules of its neighbors by listening to the sync frames.

In terms of energy consumption, S-MAC outperforms MACA and PAMAS, however it has higher latencies and for improved performance in terms of low-latency some enhanced versions of S-MAC like WiseMAC, TMAC and DMAC are used [27].

## 3 CONCLUSION & FUTURE WORK

Energy efficiency in WSNs has been a hot issue and various solutions have been proposed by researchers to make an efficient utilization of the limited energy resource of these tiny sensors. Essentially, the efforts in this regard are targeted to propose energy aware protocols both at network layer and data-link layer (specifically media-access sub-layer) of the OSI model because it is the communication module that consumes most of the sensor energy. In this paper, we have summarized the research work that has been carried out both in routing domain and in MAC mechanisms in pursuit of achieving efficient utilization of energy resource. In routing domain, we have discussed the two major categories of routing protocols namely data-centric routing (protocols in this category employ data-aggregation and sometimes caching mechanisms both to reduce the number of transmissions and amount of data to sink) and hierarchical routing (protocols in this category divides the whole network into clusters where cluster-heads gather data from field sensors, perform aggregation and then transmit aggregated data to the sink). Following the energy aware routing mechanism, we then discussed how energy conservation can be achieved by MAC protocols whose objective in WSNs is to reduce energy waste caused by collisions, idle-listening, overhearing and excessive overhead. Essentially, we discussed the two categories of MAC protocols i.e., schedule-based MAC protocols (which is a class of deterministic MAC layer protocols where access to the wireless channel is pre-allocated according to a schedule and thereby they avoid contention and thus collisions) and contention-based MAC protocols (protocols in this category do not follow a schedule but rather they employ an on-demand channel-access mechanism). Effectively, contention-based MAC protocols try to minimize rather than completely avoid the chances of collisions. Although considerable improvements have been made in terms of efficient utilization of the sensor's energy resource but this comes at the expense of incurring some latency in the data transit.

Compared to wired networks and other wireless networks such as cellular networks, WSNs are still at an early stage of development and a lot of work needs to be done in order to take them to the maturity level. Along with further optimization of the communication protocols, in future we might see the use of solar panels for trickle-recharging sensor node batteries.

### References

[1]   F.L. Lewis, "Wireless Sensor Networks", in Smart Environments: Technologies, Protocols, Applications, ed. D.J. Cook and S.K. Das, Wiley, New York, 2004.

[2]   K. Akkaya and M. Younis, "A Survey of Routing Protocols in Wireless Sensor Networks", in the Elsevier Ad Hoc Network Journal, Vol. 3/3 pp. 325-349, 2005.

[3] C. Intanagonwiwat, R. Govindan and D. Estrin, "Directed diffusion: A scalable and robust communication paradigm for sensor networks", in the Proceedings of the 6th Annual ACM/IEEE International Conference on Mobile Computing and Networking (MobiCom'00), Boston, MA, August 2000.

[4] B. Krishnamachari, D. Estrin, S. Wicker, "Modeling Data Centric Routing in Wireless Sensor Networks", in the Proceedings of IEEE INFOCOM, New York, NY, June 2002.

[5] S. Lindsey and C. S. Raghavendra, "PEGASIS: Power Efficient GAthering in Sensor Information Systems", in the Proceedings of the IEEE Aerospace Conference, Big Sky, Montana, March 2002.

[6] L. Subramanian and R. H. Katz, "An Architecture for Building Self Configurable Systems", in the Proceedings of IEEE/ACM Workshop on Mobile Ad Hoc Networking and Computing, Boston, MA, August 2000.

[7] W. Heinzelman, A. Chandrakasan, and H. Balakrishnan, "Energy-efficient communication protocol for wireless sensor networks", in the Proceeding of the Hawaii International Conference System Sciences, Hawaii, January 2000.

[8] Y. Yao and J. Gehrke, "The cougar approach to in-network query processing in sensor networks", in SIGMOD Record, September 2002.

[9] K. Akkaya and M. Younis, "An Energy-Aware QoS Routing Protocol for Wireless Sensor Networks", in the Proceedings of the IEEE Workshop on Mobile and Wireless Networks (MWN 2003), Providence, Rhode Island, May 2003.

[10] Jamal N. Al-Karaki and Ahmed E. Kamal, "Routing Techniques in Wireless Sensor Networks: A Survey".

[11] W. Heinzelman, J. Kulik, and H. Balakrishnan, "Adaptive protocols for information dissemination in wireless sensor networks", in the Proceedings of the 5$^{th}$ Annual ACM/IEEE International Conference on Mobile Computing and Networking (MobiCom'99), Seattle, WA, August 1999.

[12] S. Hedetniemi and A. Liestman, "A survey of gossiping and broadcasting in communication networks", Networks, Vol. 18, No. 4, pp. 319-349, 1988.

[13] D. Braginsky and D. Estrin, "Rumor Routing Algorithm for Sensor Networks", in the Proceedings of the First Workshop on Sensor Networks and Applications (WSNA), Atlanta, GA, October 2002.

[14] R. Shah and J. Rabaey, "Energy Aware Routing for Low Energy Ad Hoc Sensor Networks", in the Proceedings of the IEEE Wireless Communications and Networking Conference (WCNC), Orlando, FL, March 2002.

[15] S. Lindsey, C. S. Raghavendra and K. Sivalingam, "Data Gathering in Sensor Networks using the Energy*Delay Metric", in the Proceedings of the IPDPS Workshop on Issues in Wireless Networks and Mobile Computing, San Francisco, CA, April 2001.

[16] A. Manjeshwar and D. P. Agrawal, "TEEN : A Protocol for Enhanced Efficiency in Wireless Sensor Networks", in the Proceedings of the 1st International Workshop on Parallel and Distributed Computing Issues in Wireless Networks and Mobile Computing, San Francisco, CA, April 2001.

[17] A. Manjeshwar and D. P. Agrawal, "APTEEN: A Hybrid Protocol for Efficient Routing and Comprehensive Information Retrieval in Wireless Sensor Networks", in the Proceedings of the 2nd International Workshop on Parallel and Distributed Computing Issues in Wireless Networks and Mobile computing, Ft. Lauderdale, FL, April 2002.

[18] Bhaskar Krishnamachari, "Networking Wireless Sensors", Cambridge Publication.

[19] Kazem Sohrabi, Daniel Manoly, and Taieb Znati, "Wireless Sensor Networks- Technology, Protocols, and Applications", A John Wiley & Sons, Inc., Publication.

[20] K. Sohrabi, J. Gao, V. Ailawadhi, G.J. Pottie, "Protocols for self-organization of a wireless sensor network", IEEE Personal Communications, October 2000, pages 16–27.

[21] S. Dulman, L. Hoesel, T. Nieberg, P. Havinga, "Collaborative Communication Protocols for Wireless Sensor Networks", European Research on Middleware and Architectures for Complex and Embedded Systems Workshop, Pisa, Italy, April 2003.

[22] R. Kalidindi, L.Ray, R. Kannan, S. Iyengar, "Distributed Energy Aware MAC Layer Protocol For Wireless Sensor Networks", in International Conference on Wireless Networks , Las Vegas, Nevada, June 2003.

[23] P. Karn, "MACA: A New Channel Access Method for Packet Radio", Proceedings of the 9th ARRL/CRRL Amateur Radio Computer Networking Conference, September 1990.

[24] S. Singh and C. S. Raghavendra, "PAMAS – Power Aware Multi-Access Protocol with Signalling for Ad Hoc Networks", Proceedings of ACM MobiCom, October 1998.

[25] W.Yee, J. Heidemann and D. Estrin, "An energy efficient MAC protocol for sensor networks", in INFOCOMM 2002, Twenty-first Annual joint conference of the IEEE computer and Communication Societies proceedings, IEEE, vol.3, pp. 1567-1576 (June 2002).

[26] Somnath Ghosh, Prakash Veeraraghavan, Samar Singh, and Lei Zhang, "Performance of a Wireless Sensor Network MAC Protocol with a Global Sleep Schedule".

[27] Anirudha Sahoo and Prashant Baronia, "An Energy Efficient MAC in Wireless Sensor Networks to Provide Delay Guarantee''.

**Dr. Sami Halawani** has done Doctorate in Computer Science from George Mason University in 1996. He also holds two Master degrees in Computer Science from Washington University and University of Miami. He is currently working as Dean of College of Computing & Information Technology Rabigh, King Abdul Aziz University, KSA. His research areas include Image Processing, Computer Graphics, Multimedia Information Retrieval & Multimedia Networking.

**Abdul Waheed Khan** has done MSc in Digital Communications Networks with Distinction from London Metropolitan University, UK in 2008. He has done his graduation in Computer Science from University of Peshawar, Pakistan. He is currently working as Lecturer at College of Computer Sciences & Information Systems, Najran University, KSA. His current research areas include Ad Hoc Networks, Sensor Networks & Wireless Networks.